# *Exploiting holographically encoded variance to transmit labelled images through a multimode optical fiber*

Liam Collard[1,2,*], Mohammadrahim Kazemzadeh[1,*] Linda Piscopo, [1,3] Massimo De Vittorio, [1,2,3,†] Ferruccio Pisanello[1,2, †]

E-mail: liam.collard@iit.it massimo.devittorio@iit.it ferruccio.pisanello@iit.it

[1]Istituto Italiano di Tecnologia, Center for Biomolecular Nanotechnologies, Arnesano, LE 73010, Italy

[2]RAISE Ecosystem, Genova, Italy

[3]Dipartimento di Ingegneria Dell'Innovazione, Università del Salento, Lecce 73100, Italy

[*]These authors contributed equally and are co-first authors

[†]These authors jointly supervised and are co-last authors of this work

**Abstract**

Artificial intelligence has emerged as promising tool to decode a phase image transmitted through a multimode fiber (MMF) by applying deep learning techniques. By transmitting tens of thousands of images through the MMF, deep neural networks (DNNs) are capable of learning how to decipher the seemingly random output speckle patterns and unveil the intrinsic input-output relationship. High fidelity reconstruction is obtained for datasets with a large degree of homogeneity, which underutilizes the capacity of the combined MMF-DNN system. Here, we show that holographic modulation can be employed to encode an additional layer of variance on the output speckle pattern, improving the overall transmissive capabilities of the system. Operatively we have implemented this by adding a holographic label to the original dataset and injecting the resulting phase image into the fiber facet through a Fourier transform lens. The resulting speckle pattern dataset can be clustered primarily by holographic label (rather than the image data), and can be reconstructed without loss of fidelity. As an application, we describe how colour images may be segmented into RGB components and each colour component may then be labelled by distinct hologram. A UNET architecture was then used to decode each class of speckle patterns and reconstruct the colour image without the need for temporal synchronisation between sender and receiver.

## Introduction

The success of wavefront shaping (WS) techniques[1] has enabled exploitation of modal diversity in multimode optical fibers (MMFs) to finely control light transmission by operating phase-only modulation at the fiber input, overcoming the inherent optical turbidity of the waveguide. WS methods to control light transmission through MMFs are based on the recording of both intensity and phase of the speckle patterns at the fiber output and on the use of phase conjugation or transmission matrix method to create the desired amplitude distribution at the fiber output[2–11]. This has led to set of novel applications, including low invasiveness neural endoscopes with sub-cellular spatial resolution [12–17], far-field imaging [18], holographic optical tweezers [19], remote control of plasmonic structures [20].

Alternatively, recent works have shown the transmissive properties of MMFs may be evaluated by artificial intelligence techniques without the need for any phase measurement at the MMF's output [21,22]. By directly projecting the screen of a phase-only spatial light modulator (SLM) onto the fiber facet via a 4f system, deep learning has been established as a technique to reconstruct the SLM pattern and thus "see" the SLM screen through the fiber, and the ability to "learn" the transmission matrix of a multimode fiber is now well understood to be a promising method to transmit image data through a multimode fiber. For such a problem, the image may either be coupled by a 4f system directly projecting the image onto the fiber facet[21,22], or indirectly by focussing the image through fourier transform lens onto the fiber facet[23]. Typically, tens of thousands of images are then coupled through the fiber to train the network and several thousand are used to validate it. Since these pioneering works, significant advancements have been made in improving the training fidelity. Recently, an attention layer has been shown to reduce the requirement of a large dataset down to hundreds of images[24] or, alternatively, a simpler network architecture (hidden layer dense neural network) may reduce the training time to several minutes[25]. Utilising training data sets where the fiber is physically or thermally perturbed, image reconstruction has been demonstrated with strong resilience to the bending[26–29], and temperature changes[30]. Wang et al investigated the role of light source line width and stability on image reconstruction[31]. Going full circle, by applying an "actor" and "trainer" network model, the machine learning technique can also be used to project desired light patterns through the fiber[32]. Despite these impressive advancements, little attention has been paid to the possibility increasing the transmissive capacity by an "all optical" method.

Here, we show how a Fourier lens-based coupling can be further exploited to increase the variance of the output speckle patterns and can "holographically label the dataset". When a dataset is modulated by multiple holograms, we find that the data primarily clusters based on the holographic carrier, which in turn act as a "label" for the transmitted data. As an example, we demonstrate how a colour image can be segmented and each colour component then projected into the distinct holographic channels on the fiber facet. The developed deep learning method is then able to not only cluster the data based on input channel but also to reconstruct the image, resulting in independent, multiple, holographic, transmission channels. Previous works have largely focussed on the transmission of grayscale images with the notable exception of Caramazza et al [23], where time-division multiplexing is used to transmit colour information requiring temporal synchronisation between the sender (SLM) and the receiver (CCD). This synchronism is not necessary when holographic labels are employed, supporting the more general conclusion that introducing additional variance in the output speckles can help in better exploiting the wealth of information that can be transmitted through MMFs.

**RESULTS**

*Operating scheme*

Figure 1A shows the principle of encoding image data into a spatially varying hologram on the input facet of the fiber. We chose to couple the image data ($\phi_{data}$) through the hologram by combining a phase mask ($\phi_{holo}$) and the image data phase mask

$$\phi_{mask} = \arg(\exp(i(\phi_{data} + \phi_{holo}))).$$

We selected a blazed grating to create a spot scanning the MMF's input, by varying the pitch and rotation of the blazed grating on $\phi_{holo}$ the image data may be scanned across the fiber core. This should generate much higher variance in the output speckle patterns than modulating only the image, which we attribute to an higher level of orthogonality in the plane of the facet over the fourier plane (screen of SLM) as a basis for the modes of the fiber. $\phi_{data}$ can be mathematically reconstructed from $\phi_{mask}$ without loss of information and by taking the Fourier transform of $\phi_{mask}$ and it can also be verified that the image data is primarily encoded in the first order (as this is where the majority of amplitude lies). In the following, we use blazed grating for generating $\phi_{holo}$, however any other phase pattern that can increase the variance can be employed to implement the technique.

In the experimental implementations described in next paragraphs, the phase image $\phi_{mask}$ was injected into a MMF through the optical setup illustrated in Figure 1B. A 633 nm laser beam was expanded by a telescope and had its polarisation rotated to match the screen of a reflective, phase-only, SLM. The reflected light passed through a second telescope which de-magnifies the beam to match the back aperture of MO1, coupling light into the MMF. The zero diffraction from the SLM could be spatially filtered at the focal point of the telescope (L3/L4) by a removable razor blade, while the alignment of the hologram and MMF was monitored on a charge coupled device (CCD1). The transmission through the fiber was collected by MO2, and the MMF's output facet imaged on CCD2 to monitor the output speckle patterns. The SLM was given 150 ms to refresh when changing the pattern and the exposure time of the CCD was in total 60 ms.

Speckle patterns generated at the output of the MMF are fed into a deep, fully convolutional neural network based on the ResUnet architecture [33], having the role of catching the phase pattern $\phi_{mask}$ only based on the speckle data. ResUnet is an evolution of the U-Net architecture [34], enhancing its depth by incorporating a residual backbone [35]. This modification is crucial as it addresses issues such as accuracy degradation and vanishing gradients, allowing for more effective and stable training [36]. A comprehensive visualization of the network's architecture is presented in Figure 1C, including the number of layers and the shape of the output tensor at each stage, providing a clear reference for readers to grasp the network's structural intricacies. In terms of activation functions, all neurons in this network utilize the rectified linear unit (ReLU). This choice not only simplifies the network's computations but also aligns with the desired behaviour, making it a suitable choice for this specific application. Moreover, ReLU is preferred for its ability to achieve faster training speeds compared to other activation functions, contributing to the overall efficiency of the learning process [37].

In the following, we first identify and describe the increased variance in the speckle patterns dataset. Then we verify that the proposed holographic modulation provides state-of-the-art reconstruction quality and, in the final part of the manuscript, we provide the evidence that the additional variance can be employed as holographic label for the transmitted data. For this latter part we employed the approach to generate red, green and blue (RGB) labels and transmit segmented colour images. We employed a network architecture comprising three parallel ResUNet components. Each of these components generates an output, which corresponds to transmitted color. The final 3D tensor within each network (disregarding the fourth dimension for data batches) is then concatenated along the third dimension, resulting in another 3D tensor with a three-fold increase in depth for each channel. This augmented tensor subsequently undergoes further processing through a convolutional network based on the ResNet architecture, aligning its output with a fully realized RGB 3D tensor.

*Encoding variance through holographic coupling*

To verify that the amount of variance associated to the additional $\phi_{holo}$ can be greater than the one associated to $\phi_{data}$, 2000 handwritten digits (28 by 28 pixels) in the MNIST dataset were coupled through 35 holograms (corresponding to 35 positions on the input facet). The number of modes carried by the fiber is approximately 1495 at 633 nm and thus capable of carrying the 28 by 28 MNIST digits (784 pixels). The resultant speckle patterns were then analysed using principal component analysis (PCA). Panel A shows a scatter plot of the two highest variance principal components (PC1 and PC2), coloured by hologram number (1 to 35, left) or MNIST digit (right). From this visualization it is clear that the two component PCA allows to obtain clusters related to the 35 holograms, while there is only a very slight sub-clustering corresponding to digit classification. Although some overlap is observed between the clusters in a two-components graph (PC1 and PC2), including higher order components of the PCA could separate this. This can be confirmed by considering the cumulative variance explained by the PCA, shown in panel B. An inflection point at 70% of the cumulative variance is observed after the 34th component which corresponds to number of holograms-1. Decreasing the number of holograms moves the inflection point to left of the, confirming that the highest amount of variance is to be assigned to variance generated in the speckle pattern by $\phi_{holo}$ rather than by $\phi_{data}$.

To gain a better representation of the speckle patterns dataset, a method that express a much wider amount of variance for each component is therefore required. We therefore applied Uniform Manifold Approximation and Projection for Dimension Reduction (UMAP), exploiting non-linear expression of the dataset and well suited to represent fiber speckle patterns [28,38]. The UMAP analysis separates the data into 35 main clusters corresponding each hologram as shown in Figure 2C (left). Compared to the PCA, the hologram based clustering is extremely strong. Strikingly, the two component UMAP analysis is also capable of representing variance in the "digit" dataset (Figure 2C, right) despite being an "unsupervised" technique. Indeed, when coloured by handwritten digit, sub-clusters are observed in each of the 35 main clusters, with strong topological symmetry. Within each cluster, digits "0", "1", "3" and "6" appear to form distinct sub-clusters whereas "4", "7" and "9" and "2", "5" and "8" are overlaid.

### *Reconstruction of holographically coupled data*

Whilst the multivariate analysis above allows recovery of input holographic label without supervision, to reconstruct the data within a single cluster a supervised CNN is required. To verify that this is possible with no loss of reconstruction fidelity, we have compared the reconstruction of handwritten digits (grayscale) transmitted through the fiber by setting $\phi_{holo}$= zero or a blazed grating. In each case, 40000 digits were transmitted to train the Unet outlined in the experimental section, 5000 were used to validate and 5000 were used to test. The network was trained on the handwritten digit ($\phi_{data}$) regardless of $\phi_{holo}$ and the Unet network depicted in Figure 1C was used. Figure 3A shows reconstruction with $\phi_{holo}$= zero, with the first row displaying a subset of the original handwritten input digit from the MNIST dataset, while the second shows $\phi_{SLM}$ (which in this case is equivalent to $\phi_{data}$). The third row shows the output speckle patterns when each $\phi_{SLM}$ is set on the SLM and transmitted through the MMF. The Unet reconstruction of $\phi_{data}$ based on the output speckle patterns is shown in the final row. Figure 3B instead shows reconstruction with $\phi_{holo}$= blazed grating, the rows are organized as in panel A. In both cases, the reconstructions are extremely similar to the original data and each digit is easily identifiable. We quantified this by measuring the structural similarity index (SSIM) between the reconstructed image and the original.-For the test data, the average SSIM between the input image and CNN reconstructed was 0.95±0.03 for $\phi_{holo}$= Zero and 0.95±0.03 when $\phi_{holo}$= blazed grating, which is typical for handwritten digit reconstruction [29]. This is illustrated in the bar charts in Figure 3C and histogram in Figure 3D (only the data used to test the network is included in the statistical analysis and examples shown in Figure 3). The training loss and validation loss are shown in Figure 2C, with the training loss being overlapped for each coupling and therefore it can be determined that in both cases handwritten digits may be transmitted with high SSIM and equivalent training times. Ultimately, the reconstruction of $\phi_{data}$ is not impacted on whether an additional blazed grating is added on.

Moving towards more complex datasets, we applied the same test to the CIFAR (grayscale) dataset. Typically, it is more challenging to transmit natural scenes than handwritten digits as the data has higher variance and as such, the SSIM of the reconstruction is generally lower. As in the previous experiment, 40000 images were transmitted and used as a training set, while 5000 were used for validation and 5000 used to test. Reconstructions with $\phi_{holo}$= zero or a blazed grating are shown in panels 3A and 3B. As before, the top row shows $\phi_{data}$ the second shows $\phi_{mask}$ the third shows the output speckle patterns and the final shows the reconstruction of $\phi_{data}$ based on the output speckle. For the test data the SSIM between the input image and reconstructed was 0.67±0.11 for $\phi_{mask}$=0 coupled data and 0.67±0.11 when $\phi_{mask}$ is equal to a blazed grating (Figure 3C and D). Such values are typical for the reconstruction of CIFAR data or natural scenes through an optical fiber using resUNET type architecture [29]. Again, the result was analysed in terms of the training loss over the network construction and it was determined that for either natural scenes or handwritten digits, the quality of the reconstruction is unaffected by the coupling of the data in the Fourier plane of the SLM. Overall, the above results demonstrate that holographic labelling does not impact the reconstruction fidelity of the neural network, we now aim to explore potential applications of our technique.

### *Holographic modulation enables RGB image transmission*

Finally, we consider how the holographic label may be used to encode additional information in the data transmission. In particular, we associate three different $\phi_{Holo}$ to red, green and blue (RGB) channels of a color image. Firstly, the image was segmented into the single RGB components ($\phi_R, \phi_G, \phi_B$) and each was added to their a different holographic grating $\phi_{Holo,R}$, $\phi_{Holo,G}$ and $\phi_{Holo,B}$, selected so that the position of the projection were evenly displaced along an arc of the fiber core. That is, if $\phi_{Colour} = \phi_R + \phi_G + \phi_B$, then the colour image may be coupled through our MMF by sequentially showing $\arg(\exp(i(\phi_R + \phi_{Holo\,R})))$,

$\arg(\exp(i(\phi_G + \phi_{Holo\ G})))$ and $\arg(\exp(i(\phi_B + \phi_{Holo\ B})))$ on the screen of the SLM and recording the output speckle on CCD2. This process was repeated for approximately 25000 coloured CIFAR images (75000 distinct phase masks). Of which, 22500 coloured images were used to build the deep neural network and 2500 to validate. As described in the methods, we chose to train three separate UNETs to reconstruct each component. In principle, a single network could be trained that could also classify the input channel, however as holographic channels can be classified by straightforward PCA or UMAP analysis (Figure 2), we have focussed the network solely on the reconstruction.

The resultant reconstruction is shown in Figure 5, the top box shows the original images, the second box shows the speckle patterns from each of the 3 holographic channels and the third shows the reconstructed image. The reconstructed color images are extremely similar to the originals and it appears to be independent of color/intensity information. This was again quantified by measuring the SSIM between original and reconstructed images. CIFAR images are classified as images of airplanes, cars, birds, cats, deer, dogs, frogs, horses, ships, and trucks and the resultant SSIM of the reconstruction for each class is shown in panel 5B and the histogram across all classes in 5D. From the plot, we determine that the reconstruction quality is independent of image class. Across all classes in the validation data, the average SSIM was $0.77\pm0.09$ confirming that our technique is able to reconstruct color images with high fidelity. The training curves are shown in panel D, demonstrating that our network was optimised for image reconstruction.

Finally, to test that the data can be clustered by holographic label as in figure 2 PCA was applied to the dataset. The holographic variance of the speckle patterns is clearly visible in the scatter plot of PC2 and PC3 shown in panel 5E, confirming that the data is perfectly clustered by hologram. We remark that PC1 represented an intensity variance between the green and red/blue channels, therefore, the highest 3 PCs account for variation in holographic channel only, as confirmed by the inflection point of the cumulative variance explained data in inset of panel E.

**Discussion and conclusion**

In this work we show that the optical superimposition of additional variance in phase image transmission through MMF can enrich the wealth of information encoded in the speckle patterns at the fiber output. For this, we propose a modulation technique based on the introduction of additive holograms that generate artificial variance that supersedes the variance of the unmodulated speckle dataset. We demonstrate that this may be done with no loss of reconstruction fidelity by the CNN Unet employed to extract the original image from the speckles pattern. The holograms act as a label for the transmitted data, with multivariate analysis confirming each holographic labels clusters independently.

As an example application, we show how an RGB image may be transmitted through three holographically-labelled channels. The RGB image can be reconstructed without the need for a temporal synchronisation between the detector and the SLM (as in [23]) after the initial training step. We have used a blazed grating to induce holographic variance, however potentially any kind of holographic carrier may be used, provided it can generate greater variance in the output speckle patterns than the one contained in the dataset itself. We note that the average SSIM for the reconstructed color images is significantly higher than that of the grayscale images. In Figure 5 the average SSIM for colored images was $0.77\pm0.09$ whereas was $0.65\pm0.11$, $0.65\pm0.11$ and $0.65\pm0.11$ for each R, G and B component. Tentatively, it could be hypothesized that the increase in SSIM is due to a data fusion effect across all channels. The upper limit for the size of image (number of pixels) that may be transmitted by the fiber is the given by the number of modes supported by the fiber. However, for each $\phi_{holo}$, the guided modes are excited with variable favourability. By varying $\phi_{holo}$ previously weak modes may become strong, and thus the AI-based reconstruction can better exploit the modal content that can be transmitted through the waveguide resulting in a higher SSIM. On the other hand, for natural scenes there is similarity between each grayscale colour component $\phi_R$, $\phi_G$ and $\phi_B$ and thus the overall reconstruction is enhanced Figure 5. Therefore we envisage that potentially holographic labelling of grayscale data could also increase the resultant SSIM of the output data.

Overall, we envisage our technique has broad implications to improve the reconstruction/classification for outstanding problems in the field such as improving the reconstruction fidelity of natural scenes, transmitting larger images through optical fibers and lowering learning times and required computational power. As well as this, the technique may reveal more broad insights into holographic control of modal propagation in multimode optical fibers.

**Funding**

L.C. and M.K. contributed equally and are co-first authors of this work. M.D.V. and F.P. jointly supervised and are co-last authors of this work. L.C., M.D.V., and Fe.P. acknowledge funding from the European Union's Horizon 2020 Research and Innovation Program under Grant Agreement No. 828972. L.C., M.D.V. and F.P. acknowledge funding from the Project "RAISE (Robotics and AI for Socio-economic Empowerment)" code ECS00000035 funded by European Union – NextGenerationEU PNRR MUR - M4C2 – Investimento 1.5 - Avviso "Ecosistemi dell'Innovazione" CUP J33C22001220001. M.D.V. acknowledges funding from the U.S. National Institutes of Health (Grant No. U01NS094190). M.D.V., and F.P. acknowledge funding from the European Union's Horizon 2020 Research and Innovation Program under Grant Agreement No 101016787. M.D.V. acknowledges funding from the European Research Council under the European Union's Horizon 2020 Research and Innovation Program under Grant Agreement No. 692943. M.D.V., and F.P. acknowledge funding from the U.S. National Institutes of Health (Grant No. 1UF1NS108177-01). M,D,V and F,P acknowledge funding from European Research Council under the European Union's Horizon 2020 Research and Innovation Program under Grant Agreement No. 966674. F.P. acknowledges funding from the European Research Council under the European Union's Horizon 2020 Research and Innovation Program under Grant Agreement No. 677683.

**Competing interests:**

M.D.V. and F.P are founders and hold private equity in OptogeniX srl, a company that develops, produces and sells technologies to deliver light into the brain. OptogeniX did not fund the research described in this work.

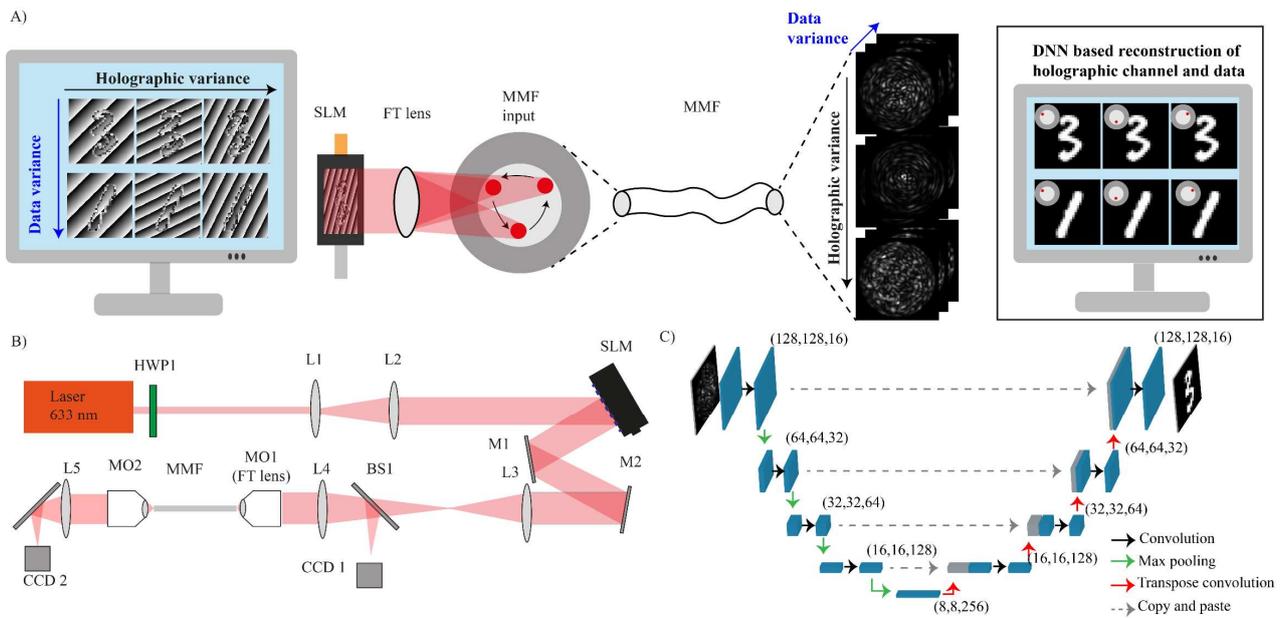

*Figure 1 – A) Principle of encoding image data in a hologram. By the addition of a blazed grating, the data is shifted around the fiber core which encodes a higher level of variance in the output speckles than the data itself. B) Optical setup used to transmit data through the MMF, M—mirror, L—lens, MO—objective, MMF—multimode fiber, SLM—spatial light modulator, BS—beam splitter, and CCD—charged coupling device. C) Structure of the implemented ResUnet grayscale image reconstruction convolutional neural networks.*

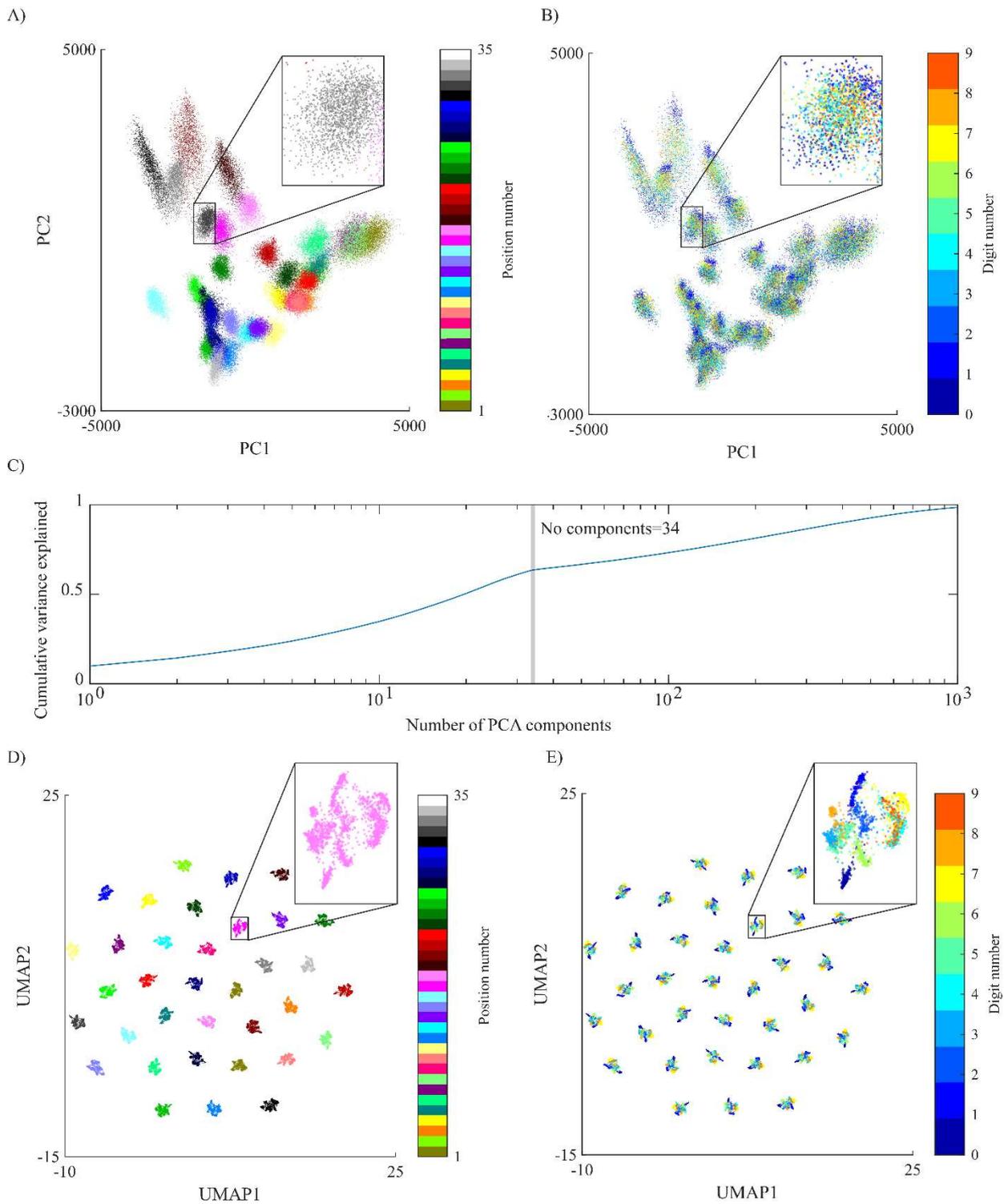

*Figure 2 – Multivariate analysis of the speckle patterns corresponding to handwritten digits coupled through 35 unqiue holograms on the input facet. A) Scatter plot of first two PCA components for all 70000 speckle patterns coloured by hologram number B) Scatter plot of first two PCA components for all 70000 speckle patterns coloured by digit number. C) Plot showing cumulative variance explained by each PC component, the inflection point at No of holograms-1 is marked. D) UMAP projection of all 70000 speckle patterns coloured by hologram number E) UMAP projection of all 70000 speckle patterns coloured by digit.*

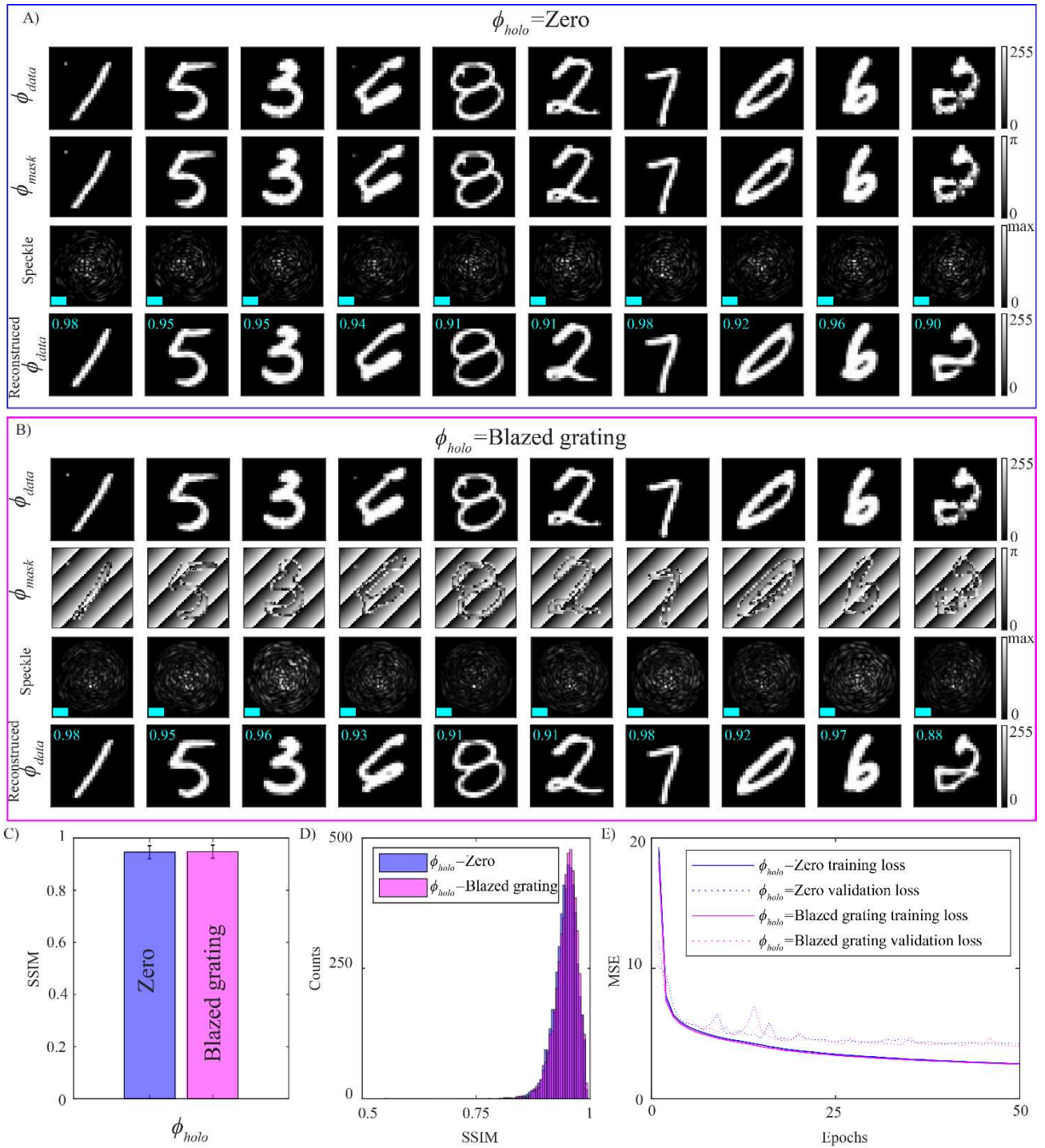

*Figure 3 - Reconstruction of greyscale handwritten digits through a MMF, coupled with either $\phi_{holo}=$ zero or a blazed grating. A) The top row are example handwritten digits to be transmitted through the fiber ($\phi_{data}$), the second row shows $\phi_{mask}$ which is shown on the SLM and in this case is equal to $\phi_{data}$ the third row shows the resultant output speckle pattern (scale bar is 10 um) and the final row shows the CNN reconstruction of $\phi_{data}$ with the SSIM inset. B) The top row are example handwritten digits to be transmitted through the fiber ($\phi_{data}$), the second row shows $\phi_{mask}$ the third row shows the resultant output speckle pattern (scale bar is 10 um) and the final row shows the CNN reconstruction of $\phi_{data}$ with the SSIM inset C) Bar charts showing the average SSIM for the test data set for each coupling D) Histograms of the SSIM for the test data set for each coupling. E) Learning curve for the development of the network for each coupling.*

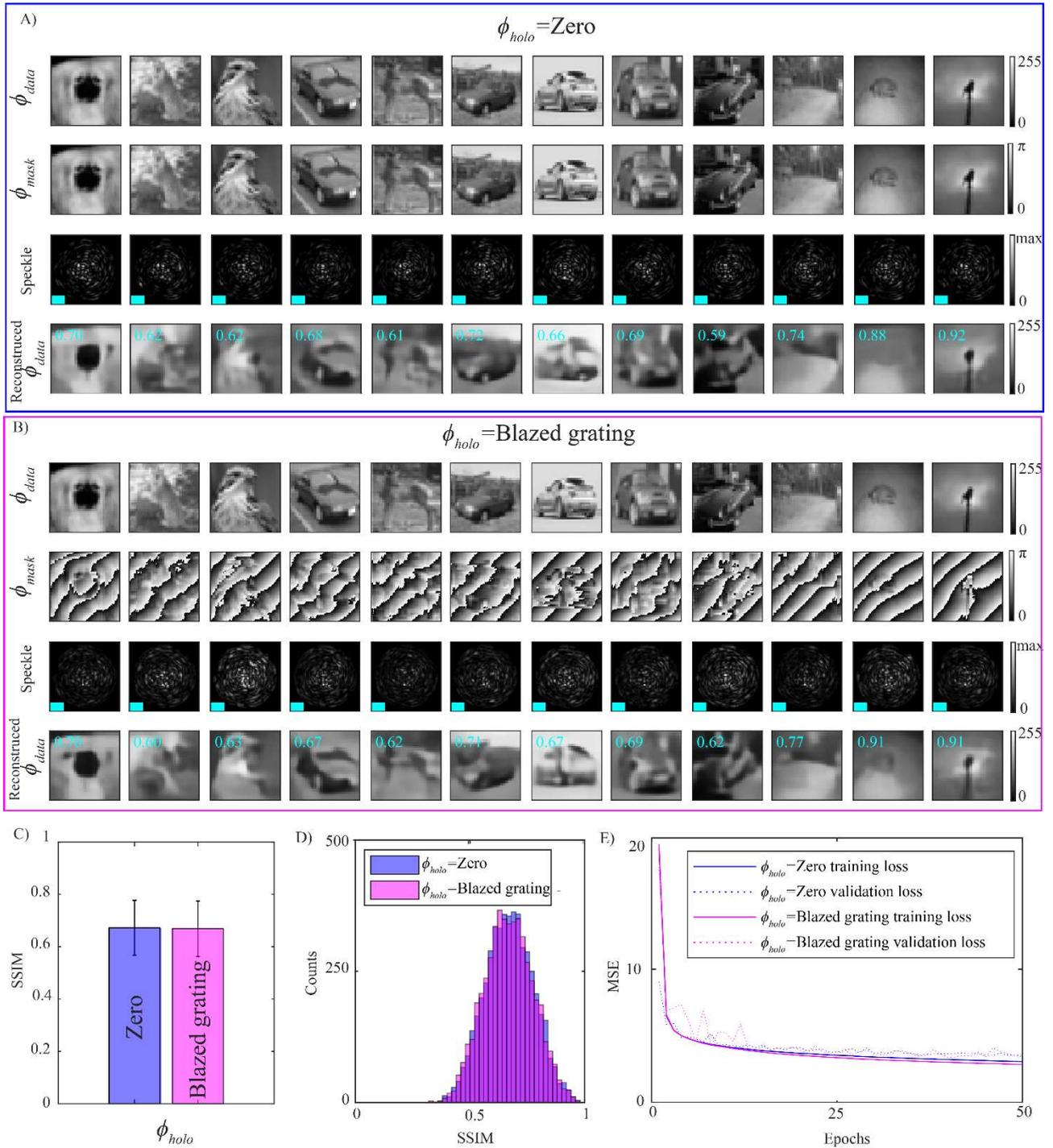

*Figure 4 - Reconstruction of greyscale natural scenes (CIFAR) coupled with either $\phi_{holo}$= zero or a blazed grating A) The top row are example handwritten digits shown on the SLM and transmitted through the fiber ($\phi_{data}$), the second row shows $\phi_{mask}$ which in this case is equal to $\phi_{data}$ the third row shows the resultant output speckle pattern (scale bar is 10 um) and the final row shows reconstruction of $\phi_{data}$ with the SSIM inset. B) The top row are example handwritten digits shown on the SLM and transmitted through the fiber ($\phi_{data}$), the second row shows $\phi_{mask}$ the third row shows the resultant output speckle pattern (scale bar is 10 um) and the final row shows reconstruction of $\phi_{data}$ with the SSIM inset C) Bar charts showing the average SSIM for the test data set for each coupling D) Histograms of the SSIM for the test data set for each coupling. E) Learning curve for the development of the network for each coupling.*

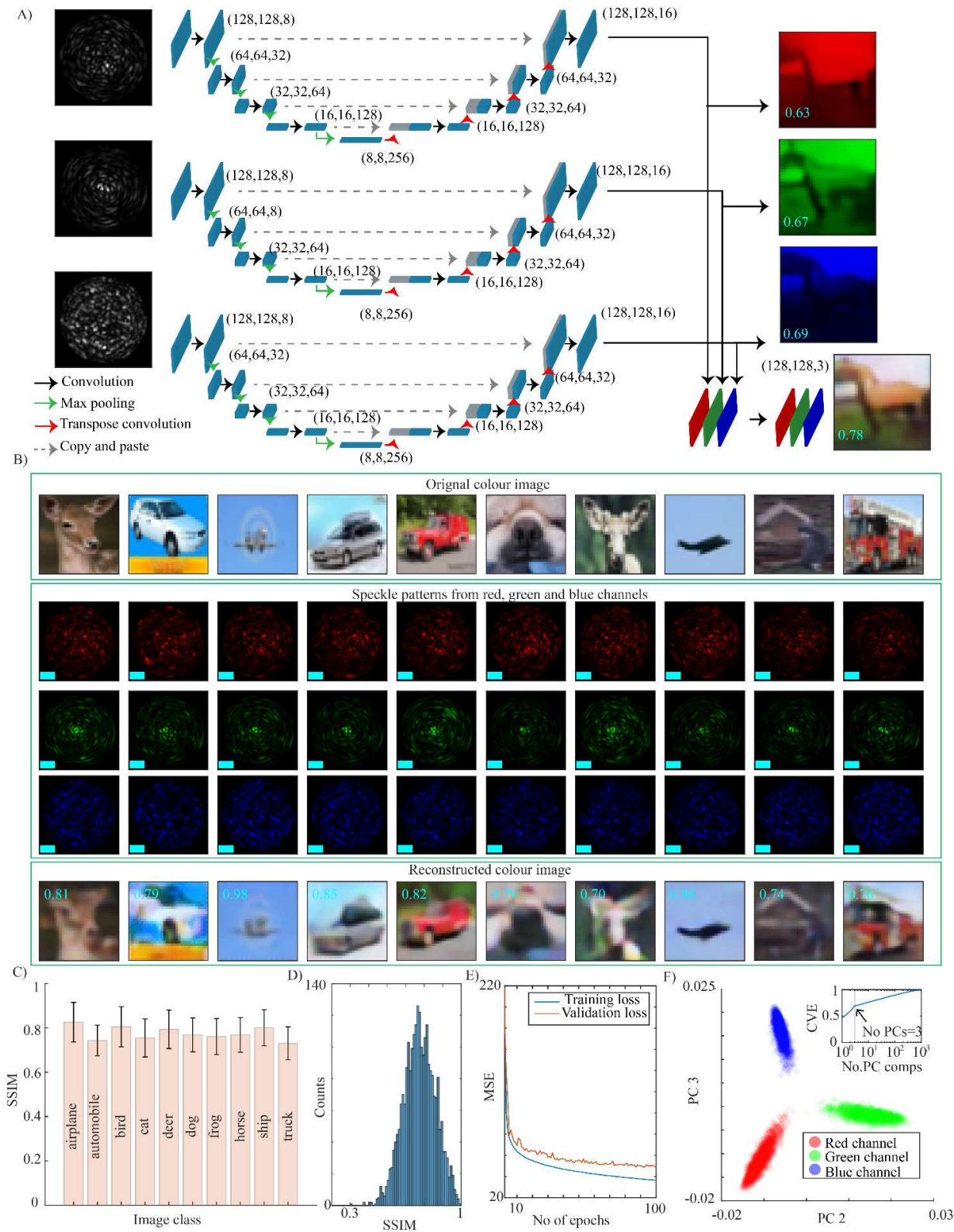

*Figure 5 – A) Structure of the implemented ResUnet type color image reconstruction convolutional neural networks. B) Reconstruction of coloured CIFAR images through multimode fiber original image (top box), corresponding three speckle patterns (middle box) and reconstructed image (bottom box) with SSIM value inset. C) Bar chart showing average SSIM for each image class (test data) D) Histogram of SSIM (test data). E) Curves showing the training and validation loss over 100 epochs, the final was used for the reconstruction shown here. F) Scatter plot of second and third PCA component for all 75000 speckle patterns, the data is perfectly clustered by holographic channel, the inset shows the cumulative variance explained by each principal component with the inflection at no of components =3 marked.*